\documentclass{PoS}

\title{Steep-spectrum sources and the duty cycle of the radio emission}

\ShortTitle{Steep-spectrum sources and the duty cycle of the radio emission}

\author{\speaker{M. Orienti}, D. Dallacasa\\
        Astronomy Department, Bologna University, Italy\\
        E-mail: \email{orienti@ira.inaf.it}}

\abstract{It is currently accepted that intrinsically compact and
  bright radio sources characterized by a convex spectrum peaking 
at frequencies ranging from 100 MHz to a few GHz are young objects. 
Following the evolutionary models, these objects would evolve into the 
population of classical radio galaxies. However, the fraction of young 
radio sources in flux density-limited samples is much larger than 
what expected from the number counts of large radio sources. 
This may suggest that for some reason a significant fraction
of young objects would never become large radio
galaxies with sizes up to a few Mpc. 
The discovery of the young radio source PKS 1518+047
characterized by an uncommonly steep spectrum confirms that the radio
emission may switch off shortly after its onset. Then the source spectrum
steepens and evolves due to energy losses. If the
interruption is not temporary, the fate of the fading sources is to
disappear at frequencies lower than those explored by current radio
telescopes. Fossils of past
activities has been recently found at pc-scale distances from newly
born radio sources, suggesting the presence of short-lived objects
with an intermittent radio emission. 
}

\FullConference{10th European VLBI Network Symposium and EVN Users Meeting: VLBI and the new generation of radio arrays\\
		September 20-24, 2010\\
		Manchester Uk}

\begin{document}

\section{Introduction}

Powerful (L$_{\rm 1.4 GHz}$ $>$ 10$^{25}$ W/Hz) and intrinsically
compact ($<$ 1$^{\prime\prime}$) extragalactic radio sources 
represent a large fraction (15--30\%) of
the radio sources selected in flux-limited catalogues. Their main
characteristic is the steep synchrotron spectrum that turns over at
frequencies between 100 MHz and a few GHz, and
interpreted as due to synchrotron-self absorption 
\cite{mo08,snellen00}, although an
additional contribution from free-free absorption (FFA) 
has been found in the most
compact sources \cite{kameno00,marr01}. 
When observed with sub-arcsecond resolution these sources usually display a
two-sided morphology with a weak core, jets and mini-lobes/hotspots,
and for this reason they were termed compact symmetric objects (CSO) by
\cite{wilkinson94}. Given their intrinsically compact size and their
morphology resembling a scaled-down version of the classical powerful
FRII \cite{fr} radio galaxies, CSOs have been interpreted as
representing an early stage in the radio source evolution. 
Decisive support to this scenario
came from the determination of both kinematic \cite{polatidis03} and
radiative \cite{mm03} ages, resulting to be about 10$^{3}$--10$^{4}$
years, i.e. much smaller than the ages (10$^{7}$--10$^{8}$ years)
estimated for classical radio
galaxies with linear sizes up to a few Mpc \cite{lara00}.  \\
In this context, it is possible to draw an evolutionary path
in which CSOs are the precursors of extended radio galaxies
\cite{phillips82}. Several
evolutionary models \cite{fanti95,snellen00} have been
developed aiming at describing how the physical properties, like
luminosity and expansion velocity change as the radio source grows. 
However, many aspects, like the excess of young radio sources in
flux-limited catalogues are not reproduced by the current models and
additional explanations must be found.\\

\section{Fading objects}

A decrease in the radio luminosity as the source grows is required by the
high fraction of young radio sources in the catalogues.
The expected number of
young objects may be determined roughly from the ratio of their typical
age and the average age of the extended sources if luminosity does not
change during the source growth. The fraction of young objects derived in this
way is a few orders of magnitude lower than what found from the source
counts. However,
in the evolutionary models the luminosity is expected to decrease by
about one order of magnitude as the source grows from a few kpc up to
Mpc scale \cite{fanti95}. However, this is again 
not enough to reproduce the source counts.\\
A possible explanation for this discrepancy has been suggested by the
distribution of the CSOs ages which peaks around 500 years
\cite{gugliu05}, indicating that a significant fraction of young
objects may be short-lived, never becoming extended radio
sources \cite{magda10}. 
However, fading radio sources are very difficult to find
due to their very steep radio spectrum that makes them
under-represented in source catalogues.
Indeed, only a few objects have been suggested as faders so far, based
on the absence of active regions \cite{magda05,magda06}, and the
distribution of spectral index found steep across the whole source,
like in the case of PKS\,1518+047 \cite{mo10}. \\

\section{The case of PKS\,1518+047}

\begin{figure}
\begin{center}
\includegraphics{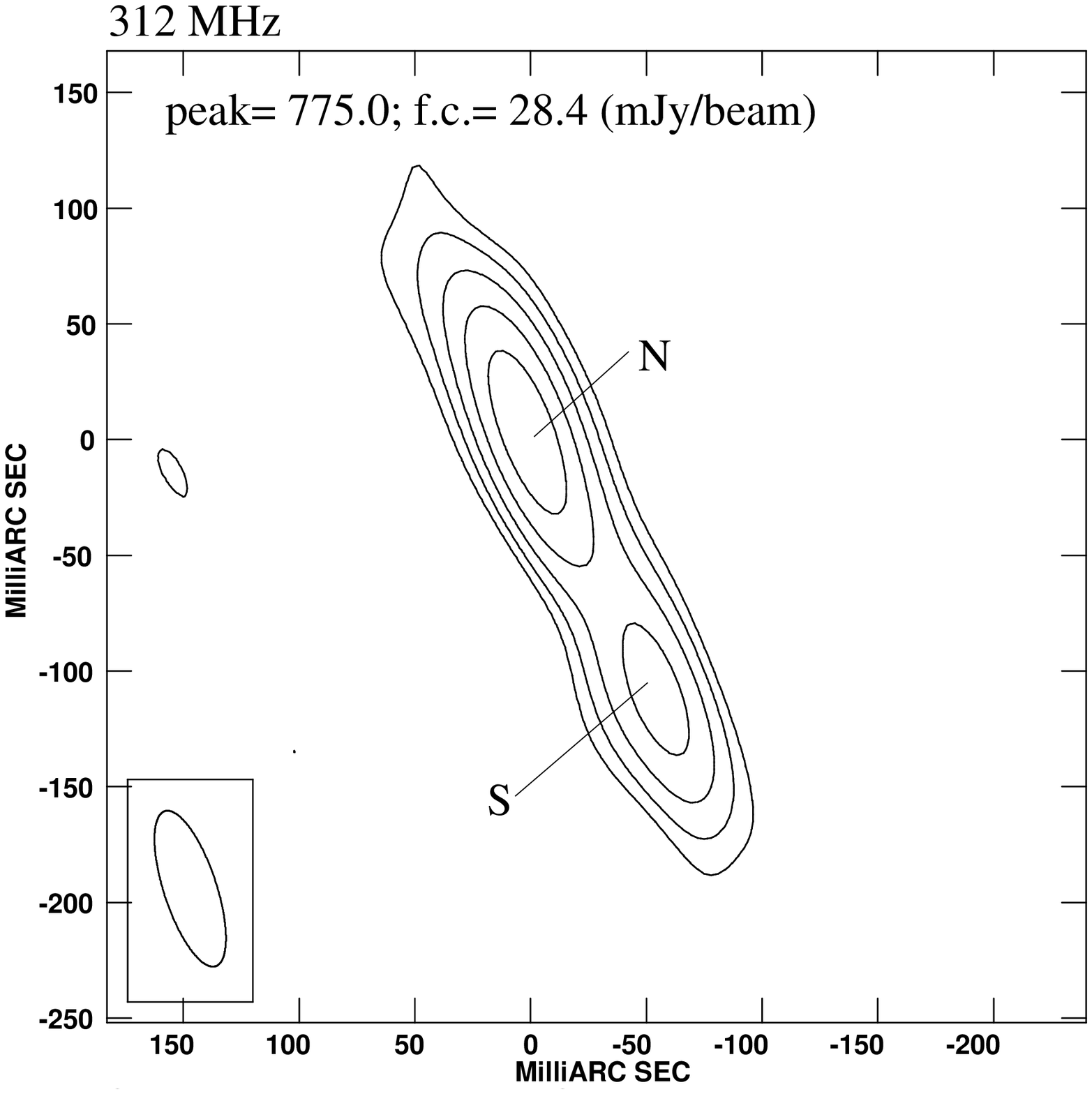}
\includegraphics{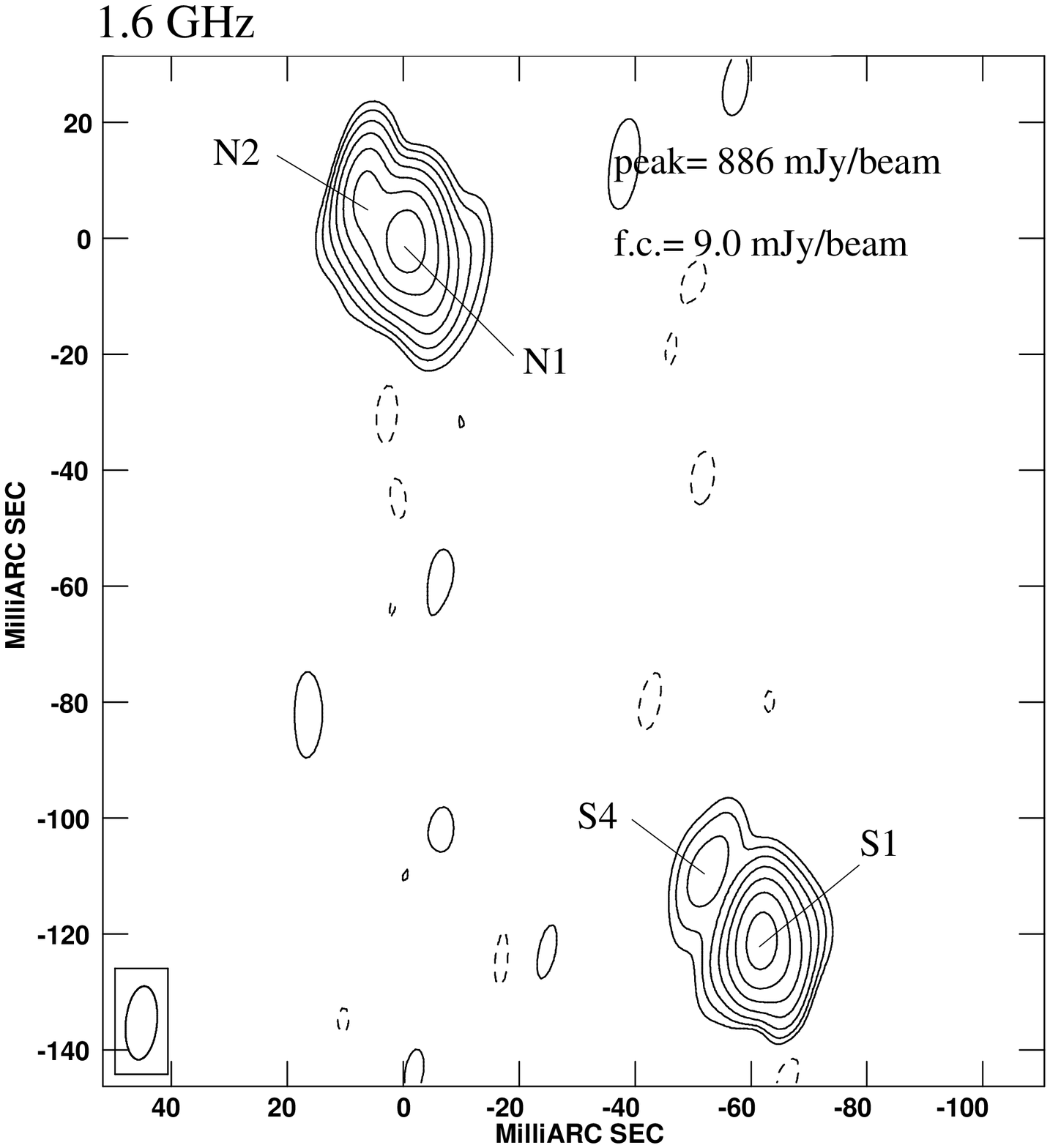}
\includegraphics{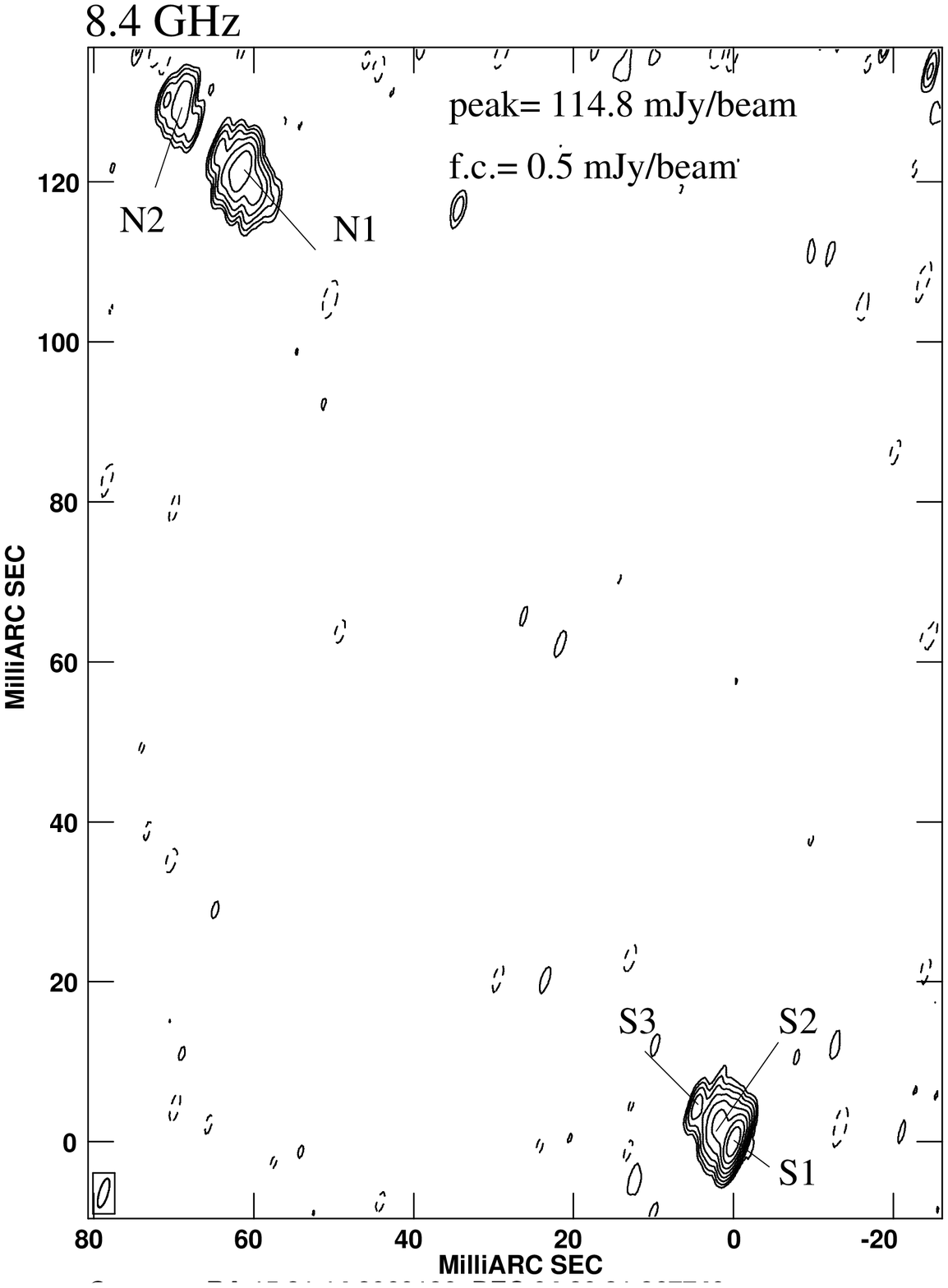}
\vspace{5.5cm}
\caption{VLBA images at 312 MHz ({\it left}), 1.6 GHz ({\it
    center}), and 8.4 GHz ({\it right}) of the fading radio source
  PKS\,1518+047. The first contour (f.c.) level corresponds to
  3 times the 1$\sigma$ noise level measured on the image. Contour
  levels increase by a factor 2. Adapted from \cite{mo10}.}
\label{1518}
\end{center}
\end{figure}

\begin{figure}
\begin{center}
\includegraphics{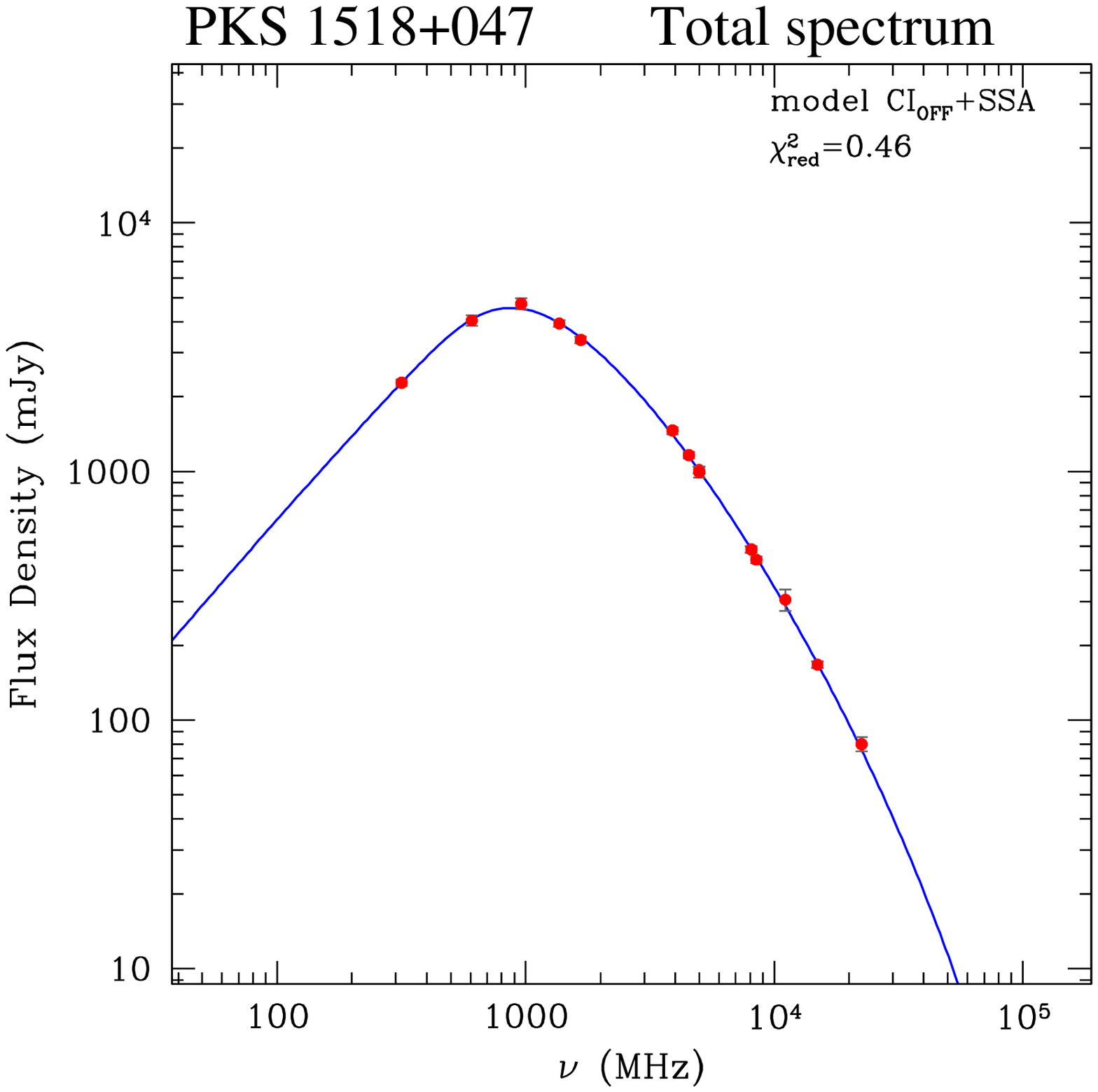}
\includegraphics{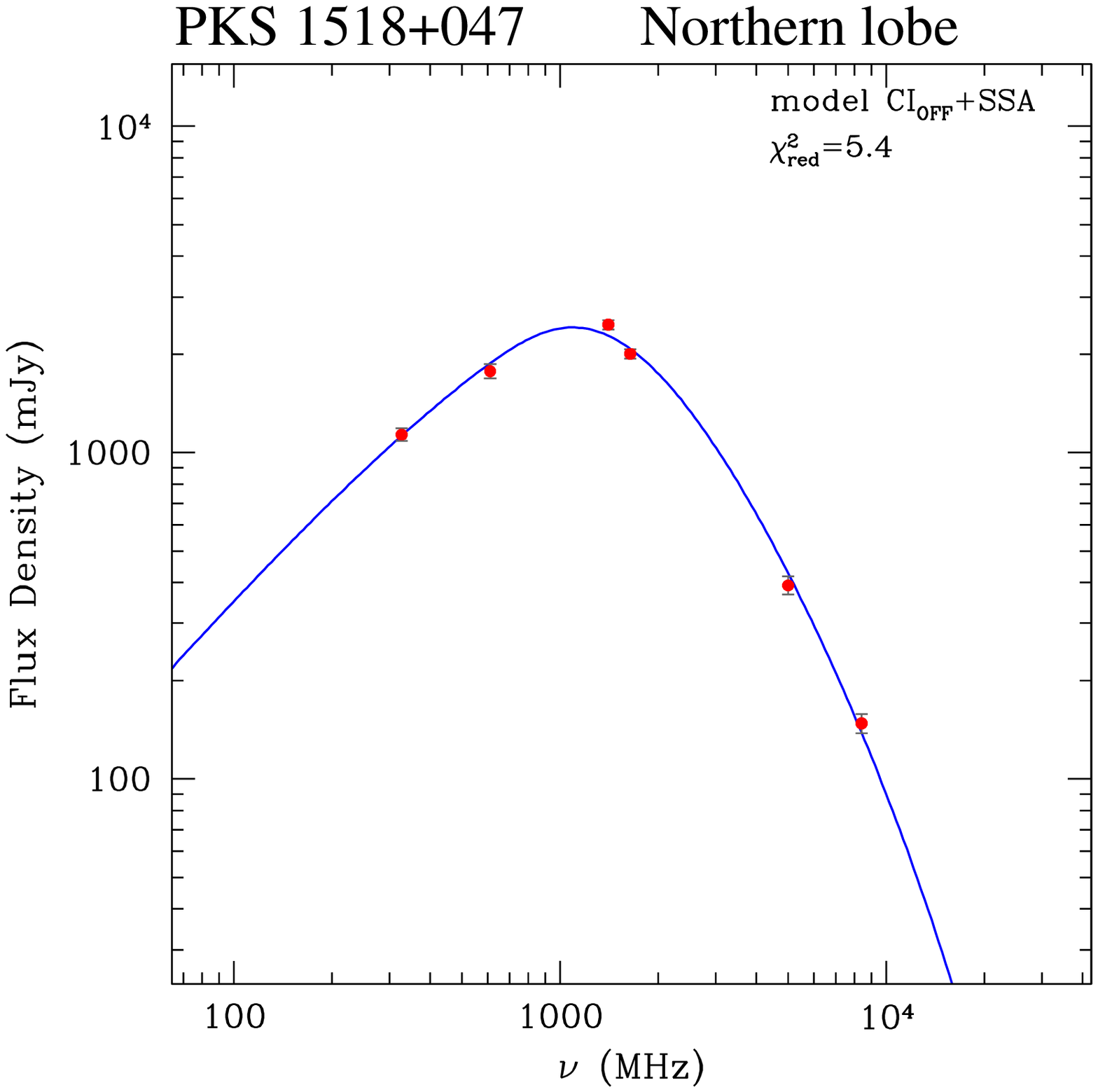}
\includegraphics{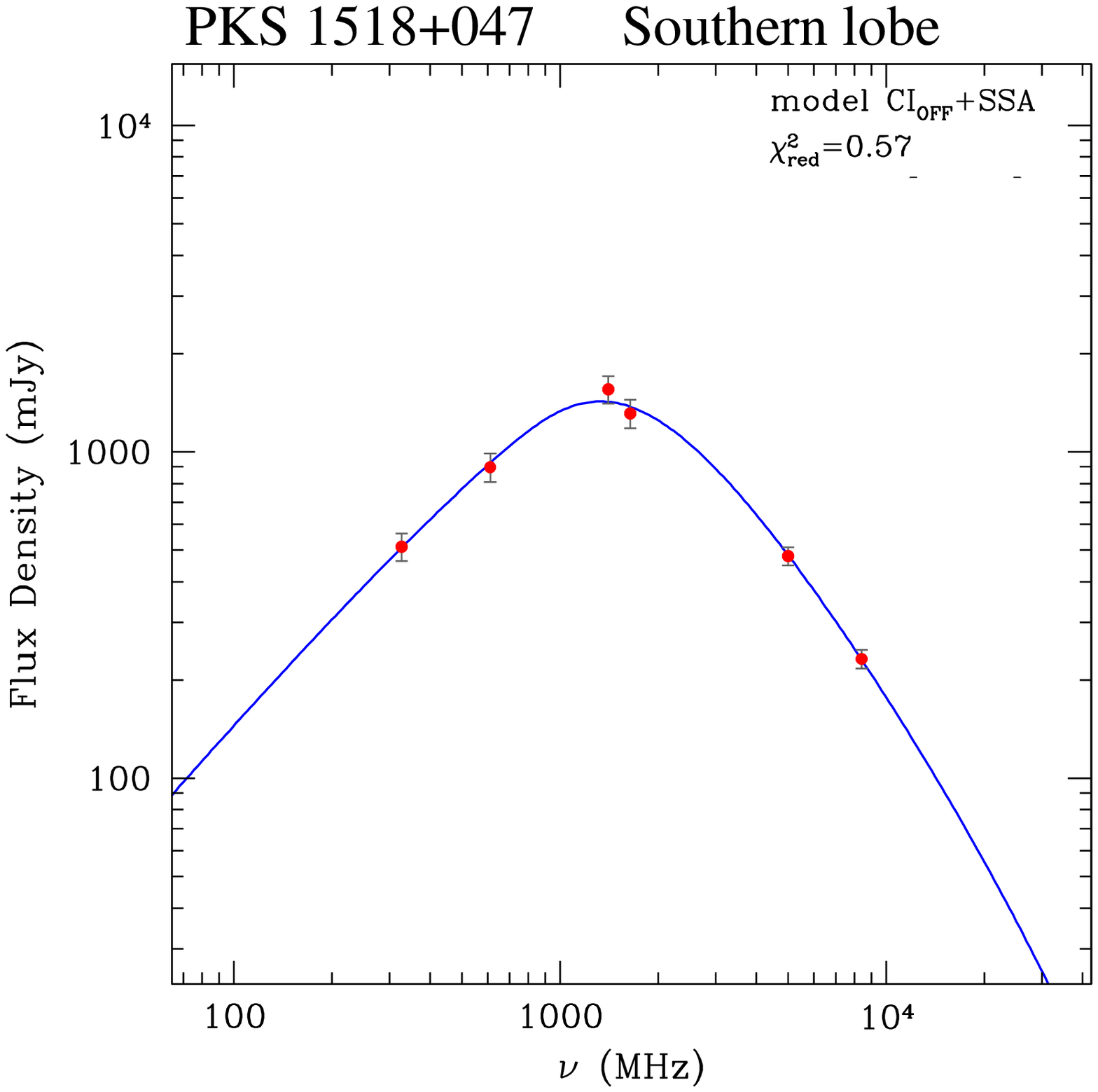}
\vspace{5cm}
\caption{The best fit to the overall spectrum of PKS\,1518+047 ({\it
    left}), and to the spectra of the northern ({\it center}) and
  southern ({\it right}) components,
  obtained using a synchrotron model in which the time spent in the
  ``fader mode'' is about 20\% of the total source age. Adapted from
  \cite{mo10}.}
\label{spettro}
\end{center}
\end{figure}

The radio source PKS\,1518+047 is a rare gem among young radio
sources. It is a powerful (L$_{\rm 1.4\,GHz}$ = 10$^{28.5}$
W/Hz) radio source 1.1 kpc in size (Fig. \ref{1518}), 
and hosted by a quasar at
$z=1.296$.  
Its radio spectrum peaks at 1 GHz and in the optically-thin
regime is uncommonly steep ($\alpha_{\rm
  4.8}^{8.4}$ = 1.2, $S \propto \nu^{- \alpha}$).\\
To understand the physical properties of this source we carried out
VLBA observations at 312, 611 and 1400/1600 MHz, and we made use of
archival VLBA data at 4.8 and 8.4 GHz to constrain both the
optically-thick and -thin part of the spectrum (a
detailed discussion on observations and data reduction can be found in
\cite{mo10}).  \\
The pc-scale resolution provided by the 312-MHz VLBA data allowed us to
resolve the source structure into two main components roughly in the
north-south direction. Both components are then further resolved in
several sub-components with
VLBA observations at higher frequencies (Fig. \ref{1518}). The
peculiarity of this source is that both the northern and southern
complexes are characterized by steep spectral indices $\alpha = 1.0 -
1.5$, indicating that no active regions, like conventional jet knots and
hotspots, are present. Strong support to the fading scenario arises
from the analysis of the synchrotron spectrum 
where injection models fail in reproducing the
spectral shape. Only models in which no particle supply is taking
place provide a good fit to the spectrum (Fig. \ref{spettro}). From
the break frequency, and assuming the equipartition magnetic field, we
compute the radiative source age that results to be 2700$\pm$600
years. On the other hand, from the best fit to spectrum we find that
the time spent by the source in the ``fader'' phase should represent
20\% ($t_{\rm OFF}=550\pm100$ years)
of the whole source lifetime, indicating that the radio emission
switched off shortly after its onset, and only electrons
with $\gamma < 600$ are still radiating \cite{mo10}. If the 
interruption of the radio activity is a temporary phase and 
the radio emission from the central engine will restart soon, 
it is possible that the source will appear again as a young radio
source, perhaps with the relics of this previous activity visible at
low frequencies. If this does not happen, 
the fate of this radio source is to emit at lower and lower
frequencies, until it disappears below the frequencies explored by
current radio telescopes.\\

\section{Recurrent activity?}

The discovery of fading objects among the population of young radio
sources may provide a interesting explanation for the excess of the source
counts. The presence of a population of short-lived objects related to
an intermittent activity of the central engine has been recently
postulated by \cite{czerny} as due to radiation
pressure instability in the accretion disk. \\
The idea that the radio emission may be a
recurrent phenomenon was suggested by \cite{baum90} after
discovering in the radio galaxy J0111+3906 an off-axis 
diffuse steep-spectrum emission at about 60 kpc away from
the newly born ($t_{\rm age} \sim 370$ years, \cite{owsianik98}), 
compact (22 pc in size) structure. In this source the
low-surface brightness feature is likely the reminiscence of
a past activity that must have lasted about 10$^{7-8}$ years in order to reach
a distance of 60 kpc from the source core. 
Recently, fossils of previous activity at parsec-scale distance from
the reborn source have been found in the two very young ($\leq$10$^{3}$
years) radio galaxies 
OQ\,208 \cite{luo07}, and J1511+0518 \cite{mo08}. Extended 
features located at pc-scale distances from the central object may be 
the relic of a far more recent previous activity that occurred 
about 10$^{3}$-10$^{4}$ years ago, suggesting that at the beginning of
the radio activity several subsequent short bursts may take place
before the development of large radio sources \cite{mo08}.\\ 
 
\section{A sample of short-lived candidates}

\begin{figure}
\begin{center}
\includegraphics{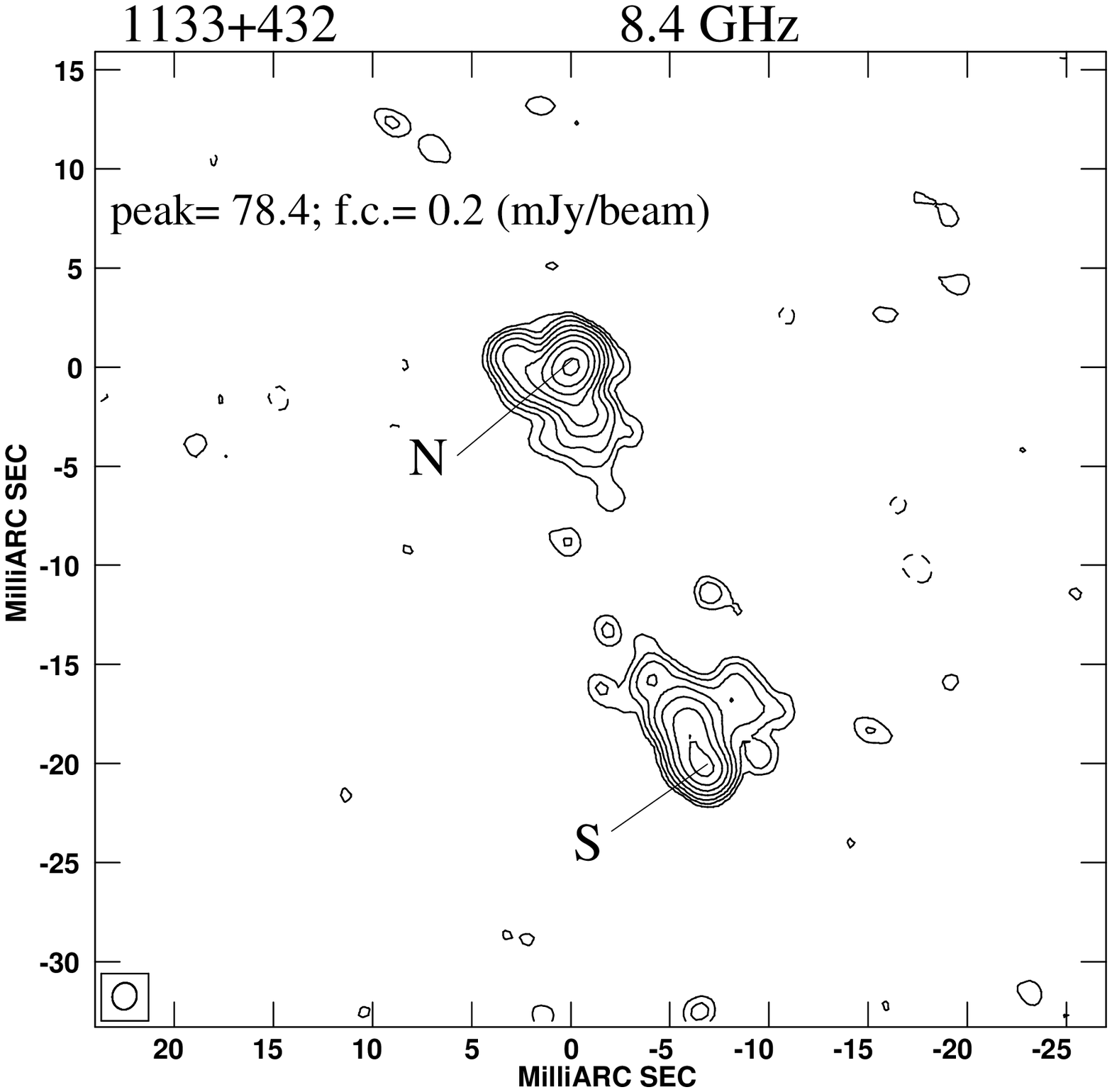}
\includegraphics{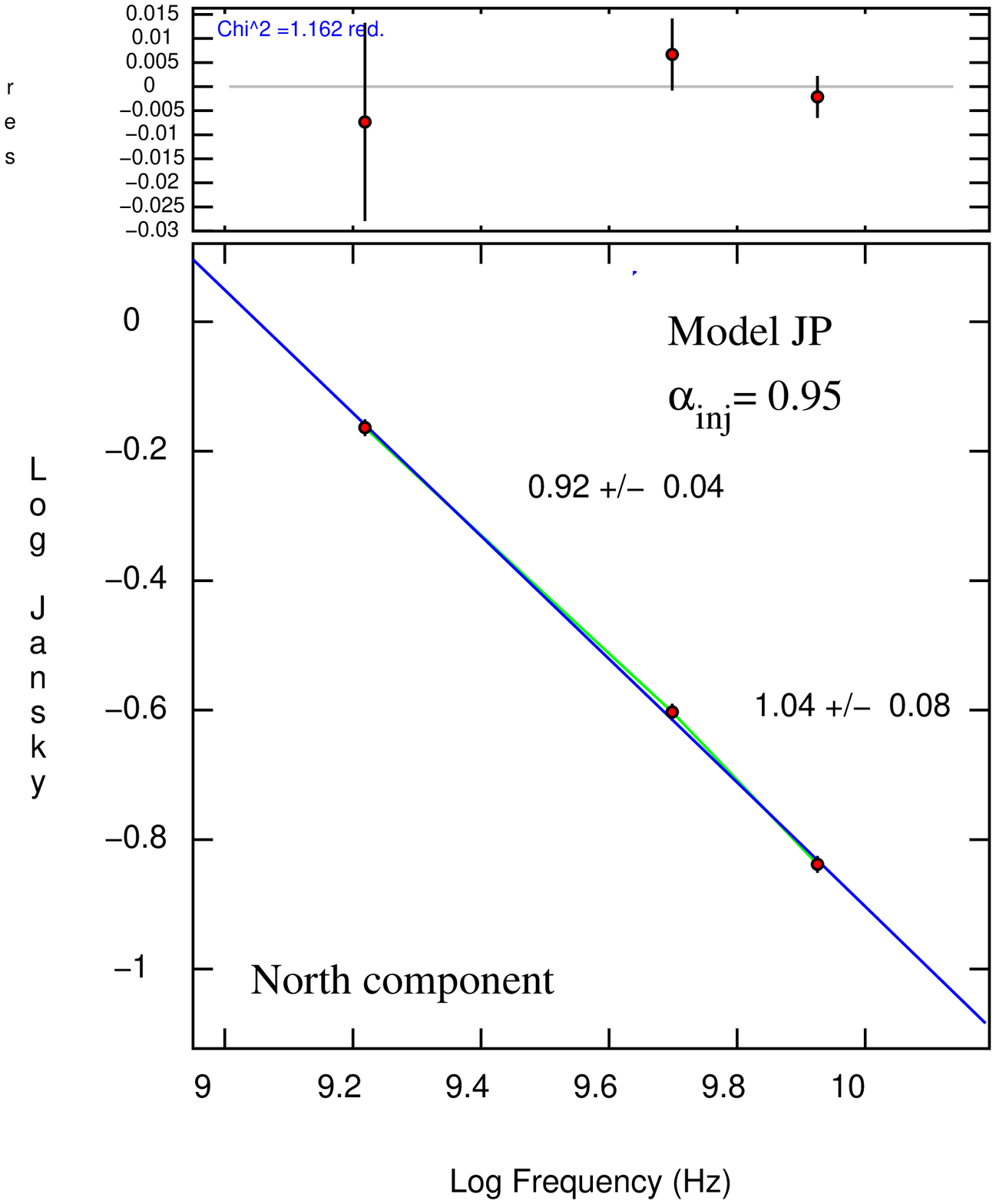}
\includegraphics{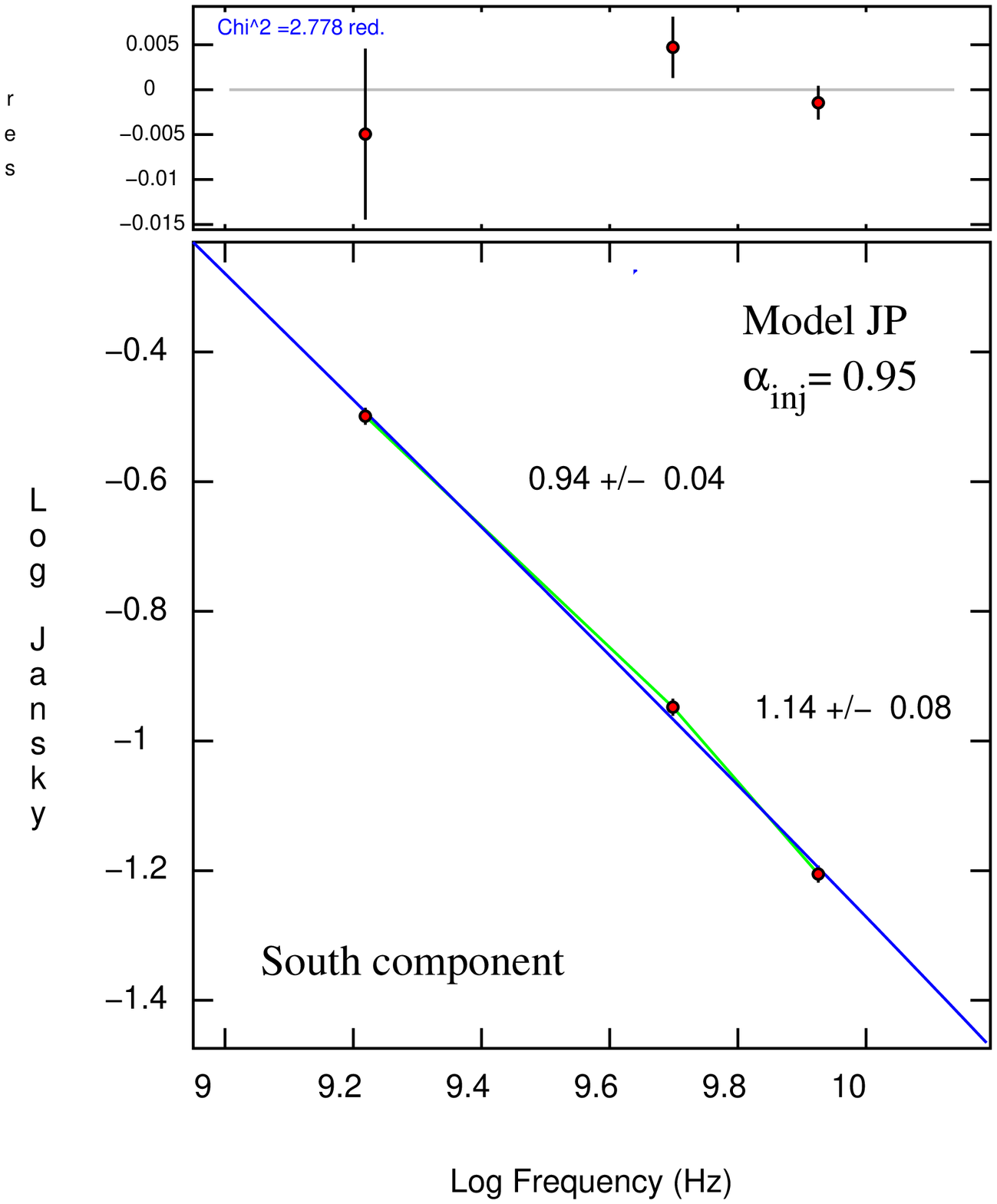}
\vspace{6cm}
\caption{The radio source B1133+432 has an example of a short-lived
  candidate. The preliminary spectral analysis does not point out the
  presence of active galaxies, like hotspots, since the spectral index
is found steep across the entire source. Adapted from \cite{mo04}.}
\label{1133}
\end{center}
\end{figure}

So far there are not statistically complete samples of short-lived
objects given their difficulty to be picked up in conventional
flux-limited radio catalogues. Furthermore, to unambiguously identify
a radio source as a short-lived object it is necessary to know the
spectral index distribution across the whole source in order to be
sure that no active regions are still present. \\
With the aim of determining the incidence of short-lived objects we
selected a sub-sample of candidate fading objects from the B3-VLA CSS
complete sample \cite{cf01} which comprises objects with linear size (and thus
ages) from 100 pc (10$^{3}$ years) and 10 kpc (10$^{5}$ years). As
short-lived candidates we selected those sources with an
optically-thin spectrum steeper than $\alpha > 1$, and without
evidence of active regions from previous multi-frequency works
\cite{mo04,rossetti06}. We ended up with 18 sources: 9 with a linear
size (LS) larger than 1 kpc, and 9 with LS $<$ 1 kpc.\\
In order to reliably constrain the spectral index distribution across
the source structure, and thus to be sure about the absence of any
active regions, we are analysing archival multifrequency VLA
data, for the sources with LS$>$1 kpc, and VLBA data for those with
LS$<$ 1 kpc. For 5 sources among the most compact ones lacking high
frequency data, we obtained new 8.4 GHz VLBA observations to complement
the frequency coverage. 
A preliminary analysis of the compact radio source
B1133+432 has not pointed out any region with flattish spectral index 
(Fig. \ref{1133}),
suggesting that we are dealing with a genuine fading short-lived
object. When the spectral information will be obtained for all the
sources we will have a complete sample made of genuine fading,
steep-spectrum objects. Their fraction will then be compared with the
radio sources in the B3-VLA CSS sample in orider to have 
a clearer picture on the incidence of genuine
short-lived objects.\\ 

\end{document}